\RequirePackage[T1]{fontenc}
\documentclass[11pt]{article}

\usepackage[height=8.85in,width=6.45in]{geometry}

\usepackage[utf8]{inputenc}
\usepackage{amsmath}
\usepackage{amssymb}
\usepackage{mathtools}
\usepackage{mathrsfs}
\numberwithin{equation}{section}
\usepackage{slashed}
\usepackage{braket}
\usepackage{graphicx}
\usepackage[svgnames]{xcolor}
\usepackage[colorlinks,citecolor=DarkGreen]{hyperref}
\usepackage{cite}
\usepackage{tikz}
\usepackage{tikz-cd}
\usepackage{times}
\usepackage{courier}
\usepackage{bm}
\usepackage{subfig}

\usepackage{xcolor}
\usepackage{mdframed}

\renewenvironment{figure}[1][]{
  \begin{originalfigure}[#1]
    \begin{mdframed}[linecolor=black!0,backgroundcolor=black!1]
}{
    \end{mdframed}
  \end{originalfigure}
}

\def\Aut{\mathrm{Aut}}
\def\avar{\mathop{\mathrm{avar}}}
\def\BC{\mathbb{BC}}
\def\ch{\mathop{\mathrm{ch}}}
\def\DS{\mathbb{DS}}
\def\Higgs{\cM_\text{\upshape Higgs}}
\def\HK{\mathcal{HK}}
\def\Hol{\mathrm{Hol}}
\def\Hom{\mathrm{Hom}}
\def\Hyp{\mathop{\mathrm{Hyp}}}
\def\IIB{\mathrm{IIB}}
\def\inst{\text{inst}}
\def\inv{\text{inv}}
\def\Ker{\mathop{\mathrm{Ker}}}
\def\MN{\mathrm{MN}}
\def\prin{\text{prin}}
\def\Rep{\mathop{\mathrm{Rep}}}
\def\SB{\mathbb{SB}}
\def\SCI{Z^\text{SCI}}
\def\SO{\mathrm{SO}}
\def\so{\mathfrak{so}}
\def\SU{\mathrm{SU}}
\def\su{\mathfrak{su}}
\def\Spin{\mathrm{Spin}}
\def\tr{\mathop{\mathrm{tr}}\nolimits}
\def\triv{\mathrm{triv}}
\def\U{\mathrm{U}}
\def\VOA{\mathbb{VOA}}

\def\bC{\mathbb{C}}
\def\bR{\mathbb{R}}
\def\bV{\mathbb{V}}
\def\bW{\mathbb{W}}
\def\bZ{\mathbb{Z}}
\def\cA{\mathcal{A}}
\def\cH{\mathcal{H}}
\def\cM{\mathcal{M}}
\def\cN{\mathcal{N}}
\def\cQ{\mathcal{Q}}
\def\cS{\mathcal{S}}
\def\cT{\mathcal{T}}
\def\cV{\mathcal{V}}
\def\cX{\mathcal{X}}

\def\fe{\mathfrak{e}}

\def\fg{\mathfrak{g}}
\def\sM{\mathscr{M}}
\def\VV{\mathsf{V}}

\def\Nequals#1{$\mathcal{N}{=}#1$}

\def\gauge{\slashed-}

\def\hypergauge{\gauge\!\!\!\gauge\!\!\!\gauge}

\def\hkq{/\!/\!/}
\def\slice#1{\wr#1}

\def\vev#1{\langle#1\rangle}

\begin{document}
\begin{flushright}
IPMU17-0169
\end{flushright}
\begin{center}

{\Large \bfseries On `categories' of quantum field theories}

\vskip .5cm
Yuji Tachikawa
\vskip .5cm

\begin{tabular}{ll}
 & Kavli Institute for the Physics and Mathematics of the Universe, \\
& University of Tokyo,  Kashiwa, Chiba 277-8583, Japan
\end{tabular}

\vskip .5cm

\end{center}

\paragraph{Abstract:} 
We give a rough description of the `categories' formed by quantum field theories.
A few recent mathematical conjectures derived from quantum field theories, some of which are now proven theorems, will be presented in this language.

\section{Introduction}
Studies of quantum field theories (QFTs) by physicists have led to
various mathematical conjectures, 
some of which are investigated fruitfully by mathematicians.
Existing mathematical formulations of QFTs do not, however, explain how these conjectures are arrived at in the first place.
It seems to the author that more properties of QFTs as perceived by physicists can be formalized in a way that a better part of the process itself of conjuring of the  conjectures become understandable to mathematicians. 

For this purpose, it seems crucial to discuss not just individual QFTs but the interrelationship among them. 
In other words, we need to discuss the `categories' formed by QFTs and possible operations in those categories.
In this note, a rough description of these `categories' will be given,
and a few recent mathematical conjectures, some of which are now proven theorems, will be phrased in this language.%
\footnote{This note is an abridged but updated version of a longer review article intended for mathematicians,
which is available on the author's webpage. 
The latest version of this manuscript will also be available on the arXiv.}

\section*{Acknowledgments}
The author thanks discussions with many fellow string theorists and helpful mathematicians.
In particular, he thanks Tomoyuki Arakawa, Hiraku Nakajima and Kazuya Yonekura, who gave valuable comments on the manuscript.
The author is partially supported  by 
JSPS KAKENHI Grant-in-Aid, Wakate-A, No.17H04837 
and Kiban-S, No.16H06335 and 
by WPI Initiative, MEXT, Japan at IPMU, the University of Tokyo.

\section{The framework}
\subsection{QFTs for $\cS$-structured manifolds}
A category of QFT exists for each fixed spacetime dimension $d$ and a structure $\cS$ on manifolds.
Here, the structure $\cS$ can be e.g.~smooth structure,  Riemannian metric,  conformal structure, spin structure, etc.
We then denote by $\cQ^d_\cS$ the category of QFT defined on $\cS$-structured manifolds of dimension $d$.
(We consider a Wick-rotated, Euclidean version of QFTs in this note.)

At the very basic level, an object $Q\in \cQ^d_\cS$ assigns:
\begin{itemize}
\item a $\bC$-vector space $\cH_Q(N)$ called the space of states to each $(d-1)$-dimensional $\cS$-structured manifold $N$  without boundary, 
\item and  the transition amplitude \begin{equation}
Z_Q(M): \cH_Q(N)\to \cH_Q(N')
\end{equation} to each $d$-dimensional $\cS$-structured manifold $M$  with the incoming boundary $N$ and outgoing boundary $N'$.
\end{itemize}
They are supposed to satisfy the standard axioms of Atiyah and Segal \cite{Atiyah,Segal}, properly modified for the structure $\cS$.
In particular, for an empty set we demand $\cH_Q(\varnothing)=\bC$,
and then if $M$ is without boundary we simply have $Z_Q(M)\in \bC$, called the partition function.

Note that a QFT $Q$ determines a functor from a suitable bordism category to the category of vector spaces.
Then $\cQ^d_\cS$ is a category formed by those functors, but as morphisms we do not choose natural transformations between functors.
We will come back to the question of morphisms in Sec.~\ref{morphisms}.

Traditionally, a QFT for smooth manifolds is called a topological QFT (despite the fact that smooth manifolds and topological manifolds can have interesting differences),
a QFT for Riemannian manifolds are simply called a QFT (without adjective), 
a QFT for conformal structure is called a conformal field theory (CFT), etc.

If an $\cS'$ structure on a manifold can be obtained by forgetting some data of an $\cS$ structure,
there is a functor $\cQ^d_{\cS'}\to \cQ^d_{\cS}$, obtained by evaluating the partition function by forgetting the additional structure on the manifold.
For example, from Riemannian manifolds we can extract a smooth manifold.
Correspondingly, a QFT for smooth manifolds can be considered as a QFT for Riemannian manifolds.
Using the traditional language, a topological QFT is an example of a QFT.

From two objects $Q_{1,2}\in \cQ^d_\cS$, one can form a product $Q_1 \times Q_2 \in \cQ^d_\cS$,
such that the partition function of $Q_1\times Q_2$ is simply $Z_{Q_1\times Q_2}(M)=Z_{Q_1}(M)Z_{Q_2}(M)$.
We always have a trivial QFT $\triv\in \cQ^d_\cS$ which is the identity of this product.
This makes $\cQ^d_\cS$ a monoidal category.

\subsection{Point operators of a QFT}
Associated to a QFT $Q\in \cQ^d_\cS$ is a vector space $\cV_Q$, called the space of point operators. 
An element of $\cV_Q$ was traditionally just called an operator of the QFT.
When $\cS$ is the Riemannian structure, the Riemannian structure with spin structure, the conformal structure,
$\cV_Q$ has an action of the rotation group $\SO(d)$, its double cover $\Spin(d)$, or the conformal group $\SO(d+1,1)$, respectively.
Let us denote by $(\cV_Q)^\inv$ the subspace invariant under these groups.
Given a $d$-dimensional manifold $M$ and point operators $\varphi_i\in (\cV_Q)^\inv$
the QFT associates a complex number we denote as \begin{equation}
Z_Q(M;\varphi_1(x_1)\varphi_2(x_2)\cdots \varphi_n(x_n)) \in \bC
\end{equation} 
for distinct points $x_i\in M$, in a way multi-linear in $\varphi_i$.
This number is called the correlation function or the $n$-point function of the theory.
We can extend this construction to arbitrary elements of $\cV_Q$ by considering a suitable bundle over $M^n$.

In the traditional axiomatic quantum field theory,
one considers $M=\bR^d$ and these are the (Euclidean version of) Wightman functions.
In a unitary theory we impose the reflection positivity.

When the structure $\cS$ is the conformal structure, there is a natural isomorphism \begin{equation}
\cV_Q \simeq \cH_Q(S^{d-1}),
\end{equation} which is called the state operator correspondence.
The action of the dilatation $x\mapsto a x\in \bR^d$ in the conformal group
 on $\cV_Q$ is usually written as $a^{-\Delta}$.
In a unitary theory  $\Delta$ is positive semidefinite and gives a grading of $\cV_Q$.
Its eigenvalues are called the scaling dimension.
Take two operators $\varphi_{1,2}\in \cV_Q$ with scaling dimension $\Delta_{1,2}$.
It can be argued that  the two-point function behaves as \begin{equation}
Z_Q(M;\varphi_1(x_1)\varphi_2(x_2))  \lesssim  \text{const.}|x_1-x_2|^{-(\Delta_1+\Delta_2)} 
\qquad\text{when $|x_1-x_2|\to 0$.}
\end{equation}
$\cV_Q$ has a structure of a certain generalized kind of an algebra.
When $d=2$, it is essentially given by the axioms of the vertex operator algebras, but with both holomorphic and antiholomorphic dependence.
It should not be too difficult to write a similar set of axioms for $d>2$.
The algebra structure is known to physicists under the name of operator product expansion (OPE) algebra.

When the structure $\cS$ is the Riemannian structure
we can introduce a filtration on $\cV_Q$ by $\bR_{\ge 0}$, still called the scaling dimension,
by demanding that the above inequality holds. 

In either case, it is a general feature of the $n$-point function that it diverges as the points approach each other. 
This in particular means that the elements of $\cV_Q$ are operator-valued distributions, i.e.~distributions which take values in the space of unbounded operators on $\cH_Q(\bR^{d-1})$. 
This makes their analysis rather complicated.
In the standard algebraic QFT approach, see e.g.~\cite{Haag},
one instead considers the net of algebras of bounded operators constructed out of $\cV_Q$.
It should also be possible to construct such a net from a QFT in our sense.
That said, in various other applications of QFTs to mathematics, $n$-point functions themselves are used.
For example, the Donaldson invariants are examples of $n$-point functions of a suitable gauge theory, as interpreted by Witten \cite{Witten:1988ze}.
Because of this, the author would like to keep $\cV_Q$ as part of the data defining a QFT.

\subsection{Deformations of a QFT }
Although we have not completely defined what a QFT is, 
it should be possible to consider a family of QFTs parameterized by an arbitrary space $\sM$.
In the category $\cQ^d_\cS$ where $\cS$ is the Riemannian structure,
a fundamental fact is that given $Q_0\in \cQ^d_\cS$,
one can construct a certain universal family of QFTs parameterized by $\sM_\text{relevant}$ such that 
\begin{itemize}
\item $0\in \sM_\text{relevant}$ corresponds to $Q_0$
\item $T_0 \sM_\text{relevant} \simeq$ the subspace of $(\cV_Q)^\inv$ whose scaling dimension is $< d$.
\end{itemize}
This family $\sM_\text{relevant}$ is called the family of relevant deformations of $Q_0$.

The idea is that, given an element $\varphi\in$ the $\SO(d)$-invariant part of $\cV_Q$,
we try to define the deformed theory $Q(\lambda\varphi)$ for a small $\lambda$ by the formula \begin{equation}
Z_{Q(\lambda\varphi)}(M) := \text{``}Z_Q(M; e^{\int_M \lambda\varphi(x) d\mu})\text{''}
\end{equation} where the right hand side is, at least in an extremely naive level, defined by expanding in $\lambda$ and writing it in terms of a sum of the $n$-point functions of $\varphi(x)$.
The singularities in the integral need to be dealt with, and the convergence of the series needs to be proven.
But physicists think that it should be possible to make sense of it 
when the scaling dimension $\Delta$ of $\varphi$ is $<d$.
The author believes that it should be possible to prove this by generalizing results already available in the study of constructive QFT.

It is also a common belief among physicists that even when $\Delta=d$, the deformation should always make sense as a formal power series in $\lambda$. 
This part should also be provable by generalizing results already available in the mathematical study of perturbative QFTs.
These deformations with $\Delta=d$ are called marginal deformations.

It also happens that for some subset of operators with $\Delta=d$ one can have an actual family, not just in the sense of formal power series.
Such deformations are called exactly marginal deformations, and are of great interest to physicists.

\subsection{$G$-symmetric QFTs}
Given a structure $\cS$ and a group $G$, we can consider a new structure $\cS\times G$ which means that the manifolds come with $\cS$ structure together with a $G$-bundle with connection.
Let us introduce a special notation $\cQ^d_\cS(G):=\cQ^d_{\cS\times G}$, an object of which is called a $G$-symmetric $\cS$-structured QFT.

Given a homomorphism $\varphi:G\to G'$, we have a functor $\varphi^*:\cQ^d_\cS(G')\to \cQ^d_\cS(G)$ defined in an obvious manner.
Similarly, given $Q_1\in \cQ^d_{\cS}(G_1)$ and $Q_2\in \cQ^d_{\cS}(G_2)$,
we have $Q_1\times Q_2 \in \cQ^d_{\cS}(G_1\times G_2)$.

We see that these categories behave like categories of spaces with $G$ action. 
In the latter case, we can sometimes construct from a space $X$ with $G\times F$ action a quotient space $X/G$ with $F$ action, if the action of $G$ is sufficiently mild. 
There is a similar construction in the categories of QFTs.
Physicists call this operation the gauging of $Q$ by $G$.

Namely, from a $G\times F$-symmetric QFT $Q\in \cQ^d_{\cS}(G\times F)$,
one can sometimes gauge $G$ and construct
$Q\gauge G \in \cQ^d_{\cS}(F)$.
The idea is to define \begin{equation}
Z_{Q\gauge G}(M,A_F) = \text{``}\int_{\cM_{G,M}} Z_Q(M,A_G,A_F) d\mu \text{''}
\end{equation} where $\cM_{G,M}$ is the space of $G$-bundles with connections on the manifold $M$, $d\mu$ is a suitable measure on it, and 
$A_G\in \cM_{G,M}$ is a specific bundle with connection.

The problem is how to make this idea precise.
When $G$ is a finite group, or when $d=1$ and $G$ is compact, there is no problem, since $\cM_{G,M}$ is finite-dimensional and there is a suitable measure.
Otherwise it is an extremely hard problem. 
Making it precise when $\cS$ is the Riemannian structure, $d=4$, $Q=\triv$, $G$ being a compact Lie group, is a big part of one of the Millennium problems \cite{JaffeWitten}.
That said, physicists share a broad consensus on the condition on $d$ and $Q$ for which the gauging by a compact Lie group $G$ makes sense.
It is generally believed that there should not be a problem when $d=2$ or $3$, that it is generically impossible when $d\ge 5$, and that a simple criterion on $Q$ is agreed upon when $d=4$.
It should also be noted that the gauged theory $Q\gauge G$, when it exists, come in a family parameterized by what is called the gauge coupling constant, with no distinguished origin in the parameter space.

The notation $\gauge G$ is not at all standard but was coined for the purpose of this note.\footnote{This symbol $\gauge$ can be typeset in LaTeX as \texttt{\$\char`\\slashed\{-\}\$} after a  \texttt{\char`\\usepackage\{slashed\}} in the preamble.}
This is supposed to give the impression that the gauging adds (`$+G$') the degrees of freedom of the gauge fields but at the same time it reduces the degrees of freedom by dividing (`$/G$') by the gauge group.

\subsection{Submanifold operators and morphisms of a QFT}
\label{morphisms}
In general, associated to a QFT $Q$, we  not only have the space of point operators $\cV_Q^0:=\cV_Q$ discussed already,
but we should also have the collection of line operators $\cV_Q^1$,
the collection of surface operators $\cV_Q^2$, \ldots, 
up to the collection of  codimension-1 operators $\cV_Q^{d-1}$.
For example, a gauge theory $Q\gauge G$ naturally has line operators labeled by a representation $R$ of $G$, such that given an embedded circle $C: S^1 \to M$ we can consider \begin{equation}
Z_{Q\gauge G}(M; R(C) ) := \text{``}\int_{\cM_{G,M}} Z_Q(M,A)  \tr\Hol_R(C) d\mu \text{''}
\end{equation}
where $\Hol$ is the holonomy of the $G$-connection $A$.
These are called the Wilson line operators by physicists.
In this case, the set of labels of Wilson lines is given by $\Rep(G)$, and forms a tensor category.

From this example and others, it is reasonable to think that for a QFT $Q$,
$\cV_Q^0$ of point operators forms a kind of algebra,
$\cV_Q^1$ of line operators forms a kind of tensor category,
$\cV_Q^2$ of surface operators forms a 2-category of some sort, \ldots,
$\cV_Q^{d-1}$ of codimension-1 operators forms a $(d-1)$ category.
The codimension-1 operators are somewhat special, since a codimension-1 locus $N$ in the spacetime $M$ can split $M$ into two disconnected regions $M_1$ and $M_2$.
Therefore, we can think of a situation where we have a QFT $Q_1$ on $M_1$, another QFT $Q_2$ on $Q_2$, and a codimension-1 operator $X$ between the two.
We consider $X$ to be a morphism from $Q_1$ to $Q_2$: $X\in \Hom(Q_1,Q_2)$,
and $\cV_Q^{d-1}=\Hom(Q,Q)$.
A codimension-2 locus can separate a codimension-1 region into two regions, supporting the morphisms $X,Y\in \Hom(Q_1,Q_2)$, respectively. 
Then such a codimension-2 operator is a morphism between morphisms,
and objects in $\cV_Q^{d-2}$ are special cases: they are morphisms between the trivial morphism in $\Hom(Q,Q)$.
This relation goes down recursively to the case of  point operators.
For topological QFTs,
the resulting categorical structure of the submanifold operators are discussed in the literature under the name of the fully-extended topological QFTs, see e.g.~\cite{Kapustin:2010ta,FreedReview,Carqueville:2016kdq}.

Note that, given a $d'$-dimensional QFT $Q'$ and a $d$-dimensional QFT $Q$ with $d'<d$,
we can tautologically consider placing $Q'$ on a dimension-$d'$ submanifold $M'\subset M$, by defining \begin{equation}
Z_Q(M; Q'(M')):=Z_Q(M) Z_{Q'}(M').
\end{equation}
In particular, take $Q$ to be the trivial QFT $\triv^{d}_\cS\in\cQ^{d}_\cS$.
We can place any $Q'\in \cQ^{d-1}_\cS$ on codimension-1 subspaces.
In other words, any $(d-1)$-dimensional QFT $Q'$ is a morphism from the trivial theory in dimension $d$ to itself: \begin{equation}
\cQ^{d-1}_\cS=\Hom(\triv^{d}_\cS,\triv^{d}_\cS)=\cV^{d-1}_{\triv^{d}_\cS}.
\end{equation}
Therefore, a full understanding of the trivial theory in $d$ dimensions in this sense entails a full understanding of all QFTs in $(d-1)$-dimensions.

\subsection{Compactifications of QFTs}
In the discussions above of the submanifold operators,
we saw that QFTs in different spacetime dimensions are intimately related.
There is also another way to relate QFTs in different dimensions.
Pick a QFT $Q\in \cQ_\cS^d$, and fix a $d'$-dimensional manifold $M'$ with $\cS$-structure.
Then, we define a $(d-d')$-dimensional QFT $Q\vev{M'}$ by demanding \begin{equation}
Z_{Q\vev{M'}}(M)=Z_Q(M\times M').
\end{equation}
This operation is called the compactification of $Q$ by $M'$ by physicists.

\subsection{Anomalous and meta QFTs}
So far we have been talking about what can be called `genuine' QFTs $Q$, where the partition function $Z_Q(M)$ takes values in $\bC$.
There are, however, many `anomalous' QFTs whose partition function does not take values in $\bC$ but only in a one-dimensional $\bC$-vector space.

To specify a $d$-dimensional anomalous QFT $\tilde Q$ with structure $\cS$, one first needs to give a rule assigning one-dimensional vector spaces to  $\cS$-structured $d$-dimensional manifolds $M$.
This can conveniently done by taking a $(d+1)$-dimensional QFT $\cA\in \cQ^{d+1}_\cS$
whose Hilbert space $\cH_{\cA}(M)$ on any $d$-dimensional manifold $M$ is one dimensional.
Such a theory $\cA$ is called invertible, and we demand  \begin{equation}
Z_{\tilde Q}(M) \in \cH_\cA(M).
\end{equation}
The $(d+1)$-dimensional QFT $\cA$ is called the anomaly of the anomalous $d$-dimensional QFT $\tilde Q$.
Equivalently,  an anomalous $d$-dimensional theory $\tilde Q$ is a morphism from a trivial $(d+1)$-dimensional theory $\triv\in \cQ^{d+1}_\cS$ to an invertible theory $\cA\in \cQ^{d+1}_\cS$,
i.e.~$\tilde Q\in\Hom(\triv^{d+1}_\cS,\cA)$.
A genuine QFT is a special case where $\cA$ is also trivial.

Once we make this generalization, it is an easy step to consider also meta QFTs in $d$-dimensions:
a meta QFT $\hat Q$ is such that its partition function $Z_{\hat Q}(M)$ takes values in a finite-dimensional Hilbert space $\cH_{\cT}(M)$ of a $(d+1)$-dimensional theory $\cT$.
One important example is the theory of conformal blocks of affine Lie algebras, for which $\cT$ is the 3d Chern-Simons theory;
another is the 6d \Nequals{(2,0)} superconformal theories which will be discussed below.
Meta QFTs are called relative QFTs by mathematicians \cite{Freed:2012bs}.

\subsection{Supersymmetric QFTs}
Mathematical conjectures often arose from the study of supersymmetric QFTs, in various dimensions.
In the framework of this note, a supersymmetric QFT in $d$ dimensions is a QFT for a particular structure $\cS$ extending the Riemannian structure.
Similarly, a superconformal QFT  in $d$ dimensions is a QFT for a structure $\cS$ extending the conformal structure.
For example, for $d=4$, both supersymmetric and superconformal QFTs come in four varieties, called $\cN{=}1,2,3,4$ supersymmetric QFTs and superconformal QFTs, respectively.

Unfortunately, it seems difficult to give a concise definition of what a supersymmetric structure on a manifold is, because of the following reason.
Let us first consider the case of a superconformal structure.
A QFT $Q$ with a superconformal structure would have an action of a superconformal group on its space $\cV_Q$ of point operators. 
The Lie algebra of a superconformal group is a super Lie algebra such  that its even component contains the conformal algebra $\so(d+1,1)$ and its odd component is in a spinor representation of the conformal group.
The fact that the odd component is in a spinor representation is required from the spin-statistics theorem of the unitary QFT.
One can also argue that the super Lie algebra in question is simple.
Then it is straightforward to list all possible superconformal algebras compatible with unitarity,
and one finds that the maximum possible dimension is $d=6$ \cite{Nahm:1977tg}.
Similarly, one finds that the maximum possible dimension for supersymmetric structures is $d=11$.

Therefore there is no hope of formulating the supersymmetric structures in a way analogous to the Riemannian structure such that the formulation applies to arbitrary dimensions. 
Their existence is accidental to low dimensions in an intrinsic way.

That said, for possible dimensions, $d\le 6$ for superconformal theories and $d\le 11$ for supersymmetric theories,
there are huge amount of literature on the physics side of the community 
about the superconformal/supersymmetric structures on a manifold,
under the name of $\cN$-extended supergravity in various dimensions.

\section{Examples}
Currently, there are many examples of QFTs which are known to physicists.
Broadly speaking, there are three methods of constructions, with overlapping range of applicabilities.
Let us examine them in turn.

\subsection{Honest constructions}
One is to construct the required data so that they satisfy the axioms.
This is the only mathematically precise method at present.
It should be mentioned that even in  this case we do not usually understand the full set of submanifold operators.

\paragraph{Topological theories:}
Many topological QFTs have been constructed in this manner.
2d topological QFTs are famously equivalent to Frobenius algebras.
3d Chern-Simons theories for a compact group $G$ can be rigorously constructed 
using the Turaev-Viro and Reshetikhin-Turaev constructions.
There are 4d topological QFTs as constructed by Crane and Yetter, etc. 

\paragraph{Conformal theories in two dimensions:}
In two dimensions, vertex operator algebras capture the local properties of the holomorphic side of a conformal field theory, and there are many mathematically rigorous discussions on them.
Their behaviors on higher-genus Riemann surfaces are governed by their conformal blocks,
which have been studied   for many rational conformal field theories and also for some irrational conformal field theories.
A full-fledged conformal field theory is obtained by consistently gluing the conformal blocks on the holomorphic side and on the anti-holomorphic side.
This aspects have also been discussed rigorously by  Schweigert, Fuchs, Runkel and their collaborators, see e.g.~\cite{Schweigert:2006af}.

\paragraph{Invertible field theories:}
Invertible field theories are invertible objects in $\cQ^d_\cS$.
From the mathematical point of view, these are the first objects one has to study in order to understand $\cQ^d_\cS$, but they got the attention of many physicists relatively recently, only in the last 10 years.
Physical studies are led by condensed-matter theorists e.g.~Kitaev, Wen and collaborators \cite{Kitaev,Wen}.
A mathematical exposition for the relativistic case can be found in e.g.~\cite{Freed:2016rqq}.

It is now known that the group of the isomorphism classes of invertible field theories in $\cQ^d_\cS(G)$, when $G$ is a finite group, is given by $E_\cS^d(BG)$, where $E_\cS^*$ is a generalized cohomology theory and $BG$ is the classifying group of $G$.
Slightly more generally, one can consider QFTs defined on  $d$-dimensional manifolds $M$ with $\cS$ structure together with a map $f:M\to X$ to a space $X$ up to homotopy.
When $\cS$ is the smooth structure, the objects in $\cQ^d_\cS[X]$ is known as homotopical sigma models and have been studied by mathematicians, see e.g.~\cite{Turaev}.
For any structure $\cS$, they should form a category $\cQ^d_\cS[X]$, and $\cQ^d_\cS(G)$ for a finite group $G$ is an example where $X=BG$.
The group of the isomorphism classes of invertible field theories in $\cQ^d_\cS[X]$ should then be given by $E_\cS^d(X)$.

\paragraph{Free theories in any dimensions:}
In any spacetime dimensions, for the structure $\cS$ being the Riemannian structure with or without spin structure, the free field theories can be constructed rigorously.
First is the free scalar field theories.
This is a functor $B_d$ from the category of $G$-vector spaces to the category $\cQ^d(G)$ of $d$-dimensional QFT with Riemannian structure with $G$ symmetry.
Pick a $G$-vector space $V$. To describe $B_d(V)$,
we take a $d$-dimensional Riemannian manifold $M$ and a $G$-bundle $P$ with connection $A$.
We then construct the associated vector bundle $V\times_G P$, whose covariant derivative we denote by $D_A$.
Then we have a Laplacian $\triangle_A$ constructed from $D_A$.
Finally, the partition function $Z_{B_d(V)}(M,A)$ is defined in terms of the eigenvalues of $\triangle_A$,
and the $n$-point functions are defined in terms of the Green function of $\triangle_A$.

Second is the free fermion theories.
This is a functor $F_d$ again from the category of $G$-vector spaces to the category $\tilde\cQ^d(G)$ of $d$-dimensional possibly-anomalous QFT with Riemannian structure, spin structure and $G$ symmetry.
The partition function and the $n$-point functions are defined in a similar manner as above, but by tensoring with the spinor bundle of $M$ and using the Dirac operator instead of the Laplacian.
This is in general an anomalous field theory with an associated anomaly $\cA(F_d(V))\in \cQ^{d+1}(G)$, whose partition function is given by the eta invariant, see \cite{Dai:1994kq}.

\subsection{Using path integrals}
Another is to use the descriptions using the path integral.
This might have been the most common method among physicists until recently.
The rough idea goes as follows. 

\paragraph{Traditional descriptions:} To construct a $d$-dimensional QFT,
we first pick a set of fields.
As an example, we first choose the spacetime dimension $d$,
a compact Lie group $G$ and its representation $R$.
We consider a $G$-bundle $P$ with connection $A$ on $M$, 
and a section $\varphi$ of the vector bundle $P\times_G R$ on $M$.
We then pick a polynomial $L$ out of these field variables and their derivatives.
As an example let us take \begin{equation}
L[A,\varphi]=\frac{1}{g^2} |F|^2 + \frac12 |D_A\varphi|^2 + \VV(\varphi)
\end{equation} where $F$ is the curvature of the $G$-connection $A$, $D_A$ is the covariant derivative with respect to $A$, and $\VV$ is a $G$-invariant polynomial on $R$, usually called the potential of the system.

Then we try to  specify the QFT using the field variables and $L$ by means of the path integral.
In this example, we try to specify a QFT $Q(G,R,\VV)\in \cQ^d$ by defining the partition function as \begin{equation}
Z_{Q(G,R,\VV)}(M) := \text{``} 
\int_{\cM_{M,G,R}} e^{-\int_M  \star L(A,\varphi)  } d\mu 
\text{''}
\end{equation} where $\cM_{M,G,R}$ is the moduli space of $G$-bundles $P$ with connections $A$  together with the section $\varphi$, $d\mu$ is an appropriate measure on it, and $\star$ is the Hodge star on $M$.

Making this construction mathematically precise is an extremely difficult problem, and forms the subject of the constructive quantum field theory. 
Despite these problems, physicists have used this ill-defined construction to uncover many properties of QFTs.
Also, physicists have put the path integral on supercomputers by discretizing the spacetime and approximating the integral by a sum, which has reproduced many experimental results to reasonable accuracy. 

\paragraph{In our language:}
In the language of this note, the theory $Q(G,R,\VV)$ above is described as follows:
we first consider a $d$-dimensional free scalar theory $B_d(R)$,
which we then gauge to construct $Q_0:=B_d(R)\gauge G$.
$\VV$ is then an element of the space of operators of $B_d(R)\gauge G$, and can be used to deform the theory from $Q_0$ to $Q_0(\VV)$.
The result of the deformation is $Q(G,R,\VV):=(B_d(R)\gauge G)(\VV)$.
The Standard Model of Particle Physics is also an example of this construction, obtained by gauging $\SU(3)\times \SU(2)\times \U(1)$ of a certain $B_4(R)\times F_4(R')$ and then by deforming it.
Theoretical and experimental high-energy physicists have spent an enormous amount of efforts to pin down what is the representations $R$ and $R'$, and also what is the precise deformation which describes our real world.
Note that this includes the brain which is reading this sentence right now.

Traditionally, most physicists  only considered QFTs of the form \begin{equation}
((B_d(R)\times F_d(R'))\gauge G)(\VV)
\end{equation}
for some $G$ vector spaces $R$, $R'$ and an element $\VV\in (\cV^0_{(B_d(R)\times F_d(R'))\gauge G})^\inv$.
As such, the aim of the constructive QFT was to make this QFT construction rigorous.

In a more modern point of view, however, not all of QFTs have this form.
Still, the gauging operation obtaining $Q\gauge G$ from $Q$,
or the deformation operation obtaining $Q(\VV)$ from $Q$, 
should make sense for $Q$, $G$ and $\VV$ satisfying appropriate conditions.
Therefore, the aim of the constructive QFT should be extended to include these more generalized constructions.

\subsection{Using String/M theory}
The final method is to construct them using string theory or M-theory.
String theory and M-theory are examples of quantum gravity theories, and fall outside of the categories of QFTs discussed in this note.

\paragraph{Quantum gravity theories:}
A quantum gravity theory is, in an extremely rough sense, a QFT where we are supposed to perform the path integral over the space of the metric, not just over the space of the connections and the sections of the associated vector bundles. 

Making sense of the preceding sentence is even more difficult than making sense of the path integral of a gauge theory as above. 
The latter is difficult but physicists believe that it should be possible to carry it out for a large number of choices of $d$, $G$, $R$, $\VV$.
The former is so difficult that physicists only know a finite number of sensible examples.
There are a few in 10 dimensions, called string theories, and a unique one in 11 dimensions, called the M theory.
They are not obtained by performing the path integral over the space of the metric.
Rather, they are found accidentally.
They are also all intimately related to each other.

In this sense, from mathematicians' point of view, they are even more ill-defined than QFTs.
Still, simply assuming their mere existence is extremely powerful,
since various QFTs can be realized and studied using string/M theory.
The status might be compared with that of  Weil cohomology theories and Grothendieck motives when they were first proposed: 
the assumption of their mere existence of these concepts allows one to explain and give a unified viewpoint on many diverse phenomena.

\paragraph{6d \Nequals{(2,0)} theories:}
An important class of QFTs constructed from string theory is the 6d \Nequals{(2,0)} theories.
We start from a 10-dimensional string theory called the type IIB string theory, which roughly speaking assigns the partition function $Z_\IIB(M)$ to a ten-dimensional manifold $M$.
Now, pick a finite subgroup $\Gamma_{G} $ of $\SU(2)$ of type $G=A_n$, $D_n$ or $E_{6,7,8}$.
We define a 6d QFT $S_G $ as $S_G  := \IIB\vev{\bC^2/\Gamma_G}$, i.e.~we define its partition function for a 6-dimensional manifold $M$ by \begin{equation}
Z_{S_G} (M)=Z_{\IIB}(M\times \bC^2/\Gamma_G).
\end{equation}
They are examples of 6d \Nequals{(2,0)} superconformal meta QFTs.
There are no known descriptions of these theories via path integrals.
There is also a free 6d \Nequals{(2,0)} theory, which can be considered as $S_{\U(1)}$. 
It is strongly believed that $S_{G}$ for $G=\U(1)$, $A_n$, $D_n$ or $E_{6,7,8}$ generate all 6d \Nequals{(2,0)} theories.

\section{Four-dimensional \Nequals2 supersymmetric theories}
\label{N=2}
Now we would like to discuss the case of 4d \Nequals2 supersymmetric and superconformal theories in more detail.
We denote the categories simply by $\cQ$ and $\cQ_c$. 
The latter is a subcategory of the former.

\subsection{Basic properties}
We first recall the overall structure in this particular case. \begin{itemize}
\item Given a compact Lie group $G$ over $\bC$, there is a category $\cQ(G)$ of 4d $\cN{=}2$ theories with $G$ symmetry. 
\item Given a homomorphism $\varphi:H\to G$, there is a functor $\varphi^*:\cQ(G)\to\cQ(H)$, satisfying expected properties.
\item There is a canonical object $\triv\in \cQ(G)$ for any $G$, which behaves naturally under the functors given above. 
\item Given $Q_1\in \cQ(G_1)$ and $Q_2\in \cQ(G_2)$, we have an operation $\times$ such that $Q_1\times Q_2\in \cQ(G_1\times G_2)$.
\item In particular, using the diagonal embedding $G\subset G\times G$, we see that for $Q_{1,2}\in \cQ(G)$ we have $Q_1\times Q_2\in \cQ(G)$. $\triv$ is the unit under this product operation. 
\item If an object $Q\in \cQ(F\times G)$ satisfies certain properties, one can form $Q\hypergauge G \in \cQ(F)$.  
It is known that $Q\hypergauge G$ is a family of \Nequals2 theories
parameterized by a neighborhood of the origin of $(\bC^\times)^n$, where $n$ equals  the number of simple factors of the Lie algebra $\fg$ of $G$.
\end{itemize}
Here we introduced an operation $Q\hypergauge G$ distinct from the operation $Q\gauge G$:
When $Q\hypergauge G$ can be formed, one can definitely also form $Q\gauge G$ but it is only guaranteed to be in the category of Riemannian QFTs, but not necessarily in the category of \Nequals2 supersymmetric theories.
Rather, one needs to define \begin{equation}
Q\hypergauge G:= ([Q\times B_d(\fg_\bC) \times F_d(\fg_\bC\oplus \fg_\bC)]\gauge G)(\VV)
\end{equation} where $\VV$ is a specially chosen deformation, to guarantee that $Q\hypergauge G$ is also \Nequals2 supersymmetric.

This is analogous to the following situation in geometry: one can consider categories $\cX(G)$ of Riemannian manifolds with $G$ action. 
Then for an $X\in \cX(F\times G)$, one can often form $X/G\in \cX(F)$.
Now, consider subcategories $\HK(G)\subset \cX(G)$ formed by hyperk\"ahler manifolds with hyperk\"ahler $G$ action.
For a $Y\in \HK(F\times G)$, we can definitely construct $Y/G$ but this is only guaranteed to be $\in \cX(F)$. 
To get an object in $\HK(G)$, one needs to perform the hyperk\"ahler quotient construction: $Y\hkq G\in \HK(F)$.
There is a deeper relationship between $\cQ(G)$ and $\HK(G)$ which will be discussed below.

Before getting there, we introduce the simplest kinds of objects in $\cQ(G)$.
Given a quaternionic vector space $V$ with hyperk\"ahler $G$ action, we define a theory of free hypermultiplets based on $V$ by the formula: \begin{equation}
\Hyp(V)=B_d(V)\times F_d(V).
\end{equation} 
They are often just called hypers, and are known to be in the subcategory $\cQ_c(G)$ of superconformal theories.

\subsection{Higgs branch functor and the slicing}
So far in this note we only talked about how to construct objects in the category of quantum field theories.
Since this is  an ill-defined category,
it is of little use to serious mathematicians.
There are also functors from these still-ill-defined categories to the well-defined categories.
The Higgs branch functor $\Higgs:\cQ(G)\to \HK(G)$ is one such example.
We also introduce an associated concept which we call `slicing'.

We only describe the Higgs branch functor at the level of objects, and we will not be able to discuss how morphisms are mapped to morphisms. 
This is due to our lack of understanding of the morphisms of $\cQ(G)$ in the first place.
The same comment applies to two other functors introduced in Sec.~\ref{other}.
We will see that even with this rudimentary understanding, we arrive at nontrivial statements.

\paragraph{The Higgs branch functor:} 
This associates to a 4d \Nequals2 supersymmetric theory $Q\in \cQ(G)$ a hyperk\"ahaler manifold $\Higgs(Q)\in \HK(G)$, with the following basic properties:
\begin{itemize}
\item $\Higgs(\Hyp(V))=V$,
\item For $Q_1 \in \cQ(G_1)$ and $Q_2\in \cQ(G_2)$ we have $\Higgs(Q_1\times Q_2)=\Higgs(Q_1)\times \Higgs(Q_2)$,
\item For $Q\in \cQ(F\times G)$, we have $\Higgs(Q\hypergauge G)=\Higgs(Q)\hkq G$
where on the left hand side we perform the gauging and on the right hand side we perform the hyperk\"ahler quotient.
\item When $Q$ is \Nequals2 superconformal, $\Higgs(Q)$ is a hyperk\"ahler cone.
\item For a family $Q$ of \Nequals2 superconformal theories, $\Higgs(Q)$ is locally constant.
\end{itemize}

\paragraph{The slicing:}
Let us now introduce the concept of the slicing. To do this, we first recall the concept of the Slodowy slice.
Consider $\fg_\bC$ and take a nilpotent element $e$ in it.
It is known that we also have elements $h,f\in \fg_\bC$ so that the triple $(e,h,f)$ defines a homomorphism from $\mathfrak{su}(2)_\bC\to \fg_\bC$ and then $\SU(2)\to G$.
We denote the commutant of this $\SU(2)$ within $G$ by $G_e$.
We  define the Slodowy slice $S_e$ at $e$ by the formula \begin{equation}
S_e:=\{e+x \mid [f,x]=0, \ x\in \fg_\bC\}.
\end{equation}
The Slodowy slice $S_e$ has a natural action by $G_e$.
In the various constructions below, the results only depend on the conjugacy class of the nilpotent element $e$. 
Therefore there are essentially finite possibilities of $e$ for a given $\fg$, labeled by the nilpotent orbits.

Now, given a hyperk\"ahler space $X\in \HK(G)$, we can define a new hyperk\"ahler space $X\slice{e}\in \HK(G_e)$, which is given as a complex manifold by the expression \begin{equation}
X\slice{e} := \mu_\bC{}^{-1} (S_e) 
\end{equation}  where $\mu_\bC:X\to \fg_\bC$ is the complex part of the moment map of the $G$ action. 
It is known that we can give a hyperk\"ahler structure. 
For $e=0$, we simply have $X\slice e=X$.
The notation $\slice{e}$ is also introduced for the sake of this exposition. It is simply chosen to vaguely suggest the letter `s'.

Now, for any $Q\in \cQ(G)$, there is a QFT procedure we call the slicing of $Q$ by $e$. 
In the physics literature it is often called the nilpotent Higgsing or the partial closure of the puncture.
This results in a theory we denote by $Q\slice{e} \in \cQ(G_e)$.
This affects the Higgs branch in the expected way: \begin{equation}
\Higgs(Q\slice{e})=\Higgs(Q)\slice{e}.
\end{equation}

\subsection{Examples}
Let us now discuss examples of 4d \Nequals2 theories.

\paragraph{4d \Nequals2 gauge theories:}
We already introduced the hypermultiplet $\Hyp(V)\in \cQ(G\times F)$ for 
a quaternionic vector space $V$ with hyperk\"ahler action of $G\times F$.
Then we can form a family of 4d \Nequals2 theories
 \begin{equation}
\Hyp(V)\hypergauge G \in \cQ(F)
\end{equation} 
if the condition is right. These are called  4d \Nequals2 gauge theories, and have been intensively studied by physicists.
We easily see that \begin{equation}
\Higgs(\Hyp(V)\hypergauge G)=V\hkq G.
\end{equation}

\paragraph{Minahan-Nemeschansky theories:}
These theories are \Nequals2 superconformal theories with $E_{6,7,8}$ symmetries
specified by a positive integer $n$: \begin{equation}
\MN(E_i,n) \in \cQ_c(E_i),\qquad i=6,7,8,
\end{equation}
constructed using a variant of the type IIB theory called the F-theory.
From this construction 
it is known that \begin{equation}
\Higgs(\MN(E_i,n))=\cM^\inst_{E_i,n}
\end{equation} where the right hand side is the centered framed instanton moduli space of the group $E_i$ on $\bR^4$ with instanton number $n$.

This means that they are not an \Nequals2 gauge theory of the form $\Hyp(V)\hypergauge G$. 
If so, we would have an equality \begin{equation}
\cM^\inst_{E_i,n} = V\hkq G
\end{equation} meaning that there is an ADHM-like description for the instanton moduli spaces of exceptional groups.
But this is almost surely impossible, since no such construction is known.

For $n=1$ these theories were first studied by Minahan and Nemeschansky \cite{Minahan:1996fg,Minahan:1996cj}
and are notable as one of the earliest examples of theories which are not gauge theories,
although they used a different argument against having a gauge theory description.
More recently, gauge theory descriptions which only manifest \Nequals1 supersymmetry have been found \cite{Gadde:2015xta}, but they are not useful at present to study its Higgs branch.

\paragraph{Class S theories:}
For this construction, we start from a 6d \Nequals{(2,0)} theory $S_G $,
and compactify it on a two-dimensional surface $C_{g,n}$ of genus $g$ with $n$ punctures.
It is known that by an appropriate trick the resulting 4d theory is \Nequals2 supersymmetric
and only depends on the complex structure of $C_{g,n}$.
We then have a family of 4d \Nequals2 theory $
S_{G,g,n}
$ parameterized by $\sM_{g,n}$,  the moduli space of Riemann surfaces of genus $g$ with $n$ punctures.
This is  the class S theory, first introduced in \cite{Gaiotto:2009we,Gaiotto:2009hg},
having the following properties:
\begin{itemize}
\item  They are $\in \cQ(G^n)$, where $G$ is the simply-connected compact Lie group of type $\fg$, so that each factor of $G$ is associated to a puncture of $C_{g,n}$.
\item In particular, the family over $\sM_{g,n}$ is such that when two points on $C_{g,n}$ are exchanged,
two factors of $G$ in $\cQ(G^n)$ are exchanged.
\item They are $\in \cQ_c(G^n)$, i.e.~superconformal, when $g=0$, $n\ge 3$, or $g=1$, $n\ge 1$, or $g\ge 2$.
\item In a neighborhood of the boundary of $\sM_{g,n}$ where the genus $g$ surface $C_{g,n}$ degenerates to a connected sum of $C_{g',n'}$ and $C_{g'',n''}$ such that $g'+g''=g$ and $n'+n''=n+2$,
we have the identification that \begin{equation}
S_{G,g,n} = (S_{G,g',n'} \times S_{G,g',n''})\hypergauge G.
\end{equation}
Here,  the gauging operation on the right hand side is performed in the following manner. 
The connected sum is performed at a puncture of $C_{g',n'}$ and another of $C_{g'',n''}$.
Accordingly we have a chosen subgroup $G$ for the first puncture 
and a chosen subgroup $G$ for the second puncture.
We then perform the gauging with respect to the diagonal subgroup of these two.
The right hand side is a family over $\sM_{g',n'}\times \sM_{g'',n''} \times U$ where $U$ is a neighborhood of the origin of $\bC^\times$, which is identified with the neighborhood of the said boundary of $\sM_{g,n}$.
\item For any $G$, we always have the principal embedding $\SU(2)\to G$ and the corresponding nilpotent element $e_\prin$. 
The commutant is trivial, $G_{e_\prin}=1$.
Then we have \begin{equation}
S_{G,g,n}\slice{e_\text{prin}}=S_{G,g,n-1}.
\end{equation} 
Namely, by slicing a puncture of a class S theory by the principal nilpotent element $e_\prin$,
we can effectively remove the puncture.
\end{itemize}

From the properties listed above, it can be seen that $S_{G,g,n}$ can be constructed from $S_{G,0,3}$. 
For this reason a special abbreviation is introduced:  $T_G:=S_{G,0,3}\in \cQ(G^3)$.
From the construction, it has a natural self-equivalence permuting three factors of $G$.

\subsection{Known overlaps among the examples}

Now let us discuss some properties of the theories $T_G$ and $S_{G,g,n}$.
First, it is known that $T_{\SU(2)}=\Hyp(V\otimes_\bC V\otimes_\bC V)$,
where $V$ is the defining representation of $\SU(2)$. 

Second, for other $G\neq \SU(2)$, no  gauge theory description is known.
Still, we can slice it at three nilpotent elements $e_{1,2,3}\in \fg_\bC$ and consider the theory $T_{G}\slice{e_1,e_2,e_3}$.
For a suitable choice of $e_{1,2,3}$, they are known to be equivalent to $\Hyp(V)$ for a suitable $V$.
Here we only discuss one example. 

Take $G=\su(N)$.
A nilpotent element in $\fg_\bC$ can be conveniently described by a partition $[n_i]$ of $N$.
We take $e=[N-1,1]$, for which $G_e=\U(1)$. 
Then we have \begin{equation}
T_{\SU(N)}\slice{e} = \Hyp(V\otimes \bar W \otimes X \oplus \bar W\otimes V \otimes \bar X)
\end{equation} where $V\simeq W \simeq \bC^N$ have actions of $\SU(V)$ and $\SU(W)$ associated to the first and the second punctures,
and $X$ is the standard one-dimensional representation of $\U(1)=G_e$ associated to the third puncture sliced by $e$.

We also know the following equivalences:
\begin{align}
\MN(E_6,N) &= T_{\SU(3N)}\slice{[N^3],[N^3],[N^3]} ,\\
\MN(E_7,N) &= T_{\SU(4N)}\slice{[2N,2N],[N^4],[N^4]} ,\\
\MN(E_8,N) &= T_{\SU(6N)}\slice{[3N,3N],[2N,2N,2N],[N^6]} .
\end{align}
Note that by construction, on the right hand side are objects  in $\cQ(\SU(3)\times\SU(3)\times \SU(3))$,
$\cQ(\SU(2)\times\SU(4)\times \SU(4))$, and 
$\cQ(\SU(2)\times\SU(3)\times \SU(6))$, 
while on the left hand side are objects in
$\cQ(E_6)$, $\cQ(E_7)$, $\cQ(E_8)$.
To write an equality, we use the homomorphism $\SU(3)^3\to (\SU(3)^3/\bZ_3) \subset E_6$, etc.

\subsection{Two other functors}
\label{other}
In this section we discuss two more functors from $\cQ_c(G)$.
One is the superconformal index functor $\SCI_{p,q,t}$ and another is the vertex operator algebra functor $\VOA$.
Applied to a family of objects in $\cQ_c(G)$, both give a locally constant result.

\paragraph{Superconformal index:}
The superconformal index functor $\SCI_{p,q,t}$ is a functor which assigns to $\cQ(G)$ a virtual representation of  $G\times (\bC^\times)^3$.
We describe it using $\bC[[p,q,t]]$-valued Weyl-invariant functions on the maximal torus $T^r\subset G$, where $(p,q,t)\in (\bC^\times)^3$ and
we also use variables $z=(z_1,\ldots,z_r)\in T^r\subset G$.
We use standard abbreviations $z^w=\prod_i z_i{}^{w_i}$ for a weight $w=(w_1,\ldots,w_r)$ of $G$.  

This functor was introduced in \cite{Gadde:2011uv}.
The essential idea was to identify a differential $d:\cV_Q\to \cV_Q$ which satisfies $d^2=0$.
Then $\SCI_{p,q,t}(Q)$ is the cohomology $H(\cV_Q,d)$.
The elliptic Gamma function $\Gamma_{p,q}(x)$ defined as follows will play an important role for this functor: \begin{equation}
\Gamma_{p,q}(x)=\prod_{m,n\ge 0} \frac{1-x^{-1}p^{m+1}q^{n+1}}{1-xp^mq^n}.
\end{equation} 

The basic properties of $\SCI_{p,q,t}$ are the following.
First, the superconformal index of a free hypermultiplet is given by \begin{equation}
Z^\text{SCI}_{p,q,t}(\Hyp(V))=\prod_{w:\text{weights of $V$}} \Gamma_{p,q}(t^{1/2}z^w).\label{hyperSCI}
\end{equation}

Second,  $\SCI_{p,q,t}(Q_1\times Q_2)=\SCI_{p,q,t}(Q_1) \SCI_{p,q,t}(Q_2)$.
Third, for $Q\in \cQ_c(F\times G)$ we have 
\begin{multline}
\SCI_{p,q,t}(Q\hypergauge G) = (\frac{1}{\Gamma_{p,q}(t) \Gamma_{p,q}'(1)})^{r}
\frac1{|W_G|}\int_{T^r} \SCI_{p,q,t}(Q) \times \\
 \prod_{\alpha:\text{roots of $G$}}
\frac{1}{\Gamma_{p,q}(z^\alpha)\Gamma_{p,q}(tz^\alpha)} \prod_{i=1}^{r}\frac{dz_i}{2\pi \sqrt{-1} z_i} 
\label{gaugeSCI}
\end{multline}
where $z\in T^r \subset G$ and $|W_G|$ is the order of the Weyl group.  
The measure appearing in \eqref{gaugeSCI} is an  elliptic generalization of the Macdonald inner product and reduces to the standard Macdonald measure when $p=0$ up to a trivial rescaling.

The slicing affects the superconformal index in the following manner.
Given a nilpotent element $e\in\fg_\bC$, recall that there are $f,h\in \fg_\bC$ such that they determine a homomorphism $\SU(2)\to G$, and $G_e$ is the commutant of the image in $G$.
We then decompose $\fg_\bC$ as \begin{equation}
\fg_\bC=\bigoplus_d V_d\otimes R_d
\end{equation} where $V_d$ is the irreducible representation of dimension $d$ of $\SU(2)$ and 
$R_d$ is a representation of $G_e$.
We then define 
\begin{equation}
K_e(z)=\prod_d \prod_{w:\text{weights of $R_d$}} \Gamma_{p,q}(t^{(d+1)/2} z^w)
\end{equation} for $z$ taking values in the maximal torus of  $G_e$.
Then we have: \begin{equation}
\SCI_{p,q,t}(Q\slice e)(z)=  K_e(z)\left[ K_0(x)^{-1}\SCI_{p,q,t}(Q)(x) \right]_{x\to zt^{h/2}} .
\end{equation}

\paragraph{Vertex operator algebra:}
Let us denote the category of vertex operator algebras with a homomorphism from an affine algebra $\hat \fg$ by $\cV(G)$.
The functor $\VOA$ associates to an object $Q\in \cQ(G)$ a vertex operator algebra $\VOA(Q)\in \cV(G)$.
This functor was introduced in \cite{Beem:2013sza}. 
The essence was to show that one can locate a nice subspace of $\cV_Q$ such that the OPE algebra structure of $\cV_Q$ induces the structure of a vertex operator algebra on it.

Here we mostly use physicists' notation for the vertex operator algebras.
For a general introduction to vertex operator algebras, see e.g.~\cite{FrenkelBen-Zvi,ArakawaProc}.
In the following, a VOA always stands for vertex operator super-algebra $\bV=\oplus_n \bV_n$,
$\bV_n = \bV_{n,+}\oplus \bV_{n,-}$. 
The part $\bV_\pm =\oplus_n \bV_{n,\pm}$   are called bosonic and fermionic, respectively,
and the subscript $n$ is the eigenvalue of $L_0$.

The basic features of this functor is as follows. 
First, for $Q=\Hyp(V)\in \cQ_c(G)$, the corresponding $\VOA(Q)$ is the symplectic boson VOA $\SB(V)$ defined in the following manner: $\SB(V)$ is generated by $\SB(V)_{1/2,+}\simeq V$, with the operator product expansion given by \begin{equation}
v(z)w(0) \simeq \frac{\vev{v,w}}z
\end{equation} for $v,w\in \SB(V)_{1/2,+}\simeq V$, where $\vev{\cdot,\cdot}$ is the symplectic pairing of the quaternionic vector space $V$.

Second, we have $\VOA(Q_1\times Q_2)=\VOA(Q_1)\otimes \VOA(Q_2)$.
Third, to describe $\VOA(Q\hypergauge G)$, we define the quotient operation in the category of vertex operator algebras.
We start from an object $\bV\in \cV(G)$. 
We introduce a ghost VOA $\BC(G)$, generated by fermionic fields $b^A$ in $\BC(G)_{1,-}$ and $c_A$ in $\BC(G)_{0,-}$ for 
$A=1,\ldots,\dim \fg$ with the OPE \begin{equation}
b^A(z) c_B(w) \sim \frac{\delta^A_B}{z-w}.
\end{equation}  This has a subalgebra $\hat\fg_{+2h^\vee(\fg)}$.  Denote by $J^A_\bV$ and $J^A_\text{ghost}$ the affine $\fg$ currents of $\bV$ and $\BC(G)$ respectively.  We define \begin{equation}
j_\text{BRST}(z)=\sum_A (c_A J^A_\bV(z) + \frac12 c_A J^A_\text{ghost}(z)). 
\end{equation} Then $d=j_\text{BRST,0}$ satisfies $d^2=0$ if the level of $\hat\fg$ in $\bV$ is $-2h^\vee(\fg)$.
We then take the subspace \begin{equation}
\bW \subset \bV\otimes \BC(G)
\end{equation} defined by \begin{equation}
\bW= \bigcap_A \Ker b^A_0 
\end{equation} where $J^A_\text{total}=J^A_{\bV}+J^A_\text{ghost}$. We can check that the differential $d$ acts within $\bW$, and finally we define \begin{equation}
\bV/G:= H(\bW,d). \label{B}
\end{equation} 
We then have the following statement:
for $Q\in \cQ_c(G)$, assume that  $Q\hypergauge G \in \cQ_c(G)$.
This implies the level of $\hat\fg$ in $\VOA(Q)$ is $-2h^\vee(\fg)$, and we have
$\VOA(Q\hypergauge G)=\VOA(Q)/G$.

Third, the slicing by a nilpotent element $e$ of an object $V\in \cV(\fg)$ is defined by the quantum Drinfeld-Sokolov reduction: $V\slice e:=\DS(V,e)$. 
Then we have  $\VOA(Q\slice e)=\VOA(Q)\slice e$.

\subsection{Relation among the functors}
\paragraph{$\VOA$ to $\Higgs$:}
For a vertex operator algebra $\bV\in \cV(\fg)$, one can construct the associated variety $\avar \bV$, which is a holomorphic symplectic variety with $G$ action \cite{Arakawa,ArakawaProc}. 
This is obtained by $\mathrm{Spec}$ of Zhu's $C_2$ algebra of the vertex algebra $\VV$.
It is believed that $\Higgs(Q) = \avar \VOA(Q)$ in general \cite{Beem:2017ooy}.

\paragraph{$\VOA$ to $\SCI_{p=0,q=t}$:}
For a vertex operator algebra $\bV\in \cV(\fg)$, we can define its character $\ch \bV$ as a $\bC[[q]]$-valued function on $G$ by the following formula: \begin{equation}
G\ni z \mapsto \ch \bV(z) = \sum_{n} q^n (\tr_{\bV_{n,+}} z - \tr_{\bV_{n,-}} z).
\end{equation} 
It is known in general that \begin{equation}
\ch\VOA(Q) = \SCI_{p=0,q=t} (Q).
\end{equation} 

\paragraph{$\Higgs$ to $\SCI_{p=q=0,t=\tau^2}$:}
For a hyperk\"ahler cone $X\in \HK_c(G)$, we can define its character $\ch X$ as a $\bC[[\tau]]$-valued function on $G$ in the following manner.
We decompose the function ring $\bC[X]$ into the graded pieces $\bC[X]_n$, where the symplectic form on $X$ is normalized to have the grade $+2$. 
We then define \begin{equation}
G\ni z \mapsto \ch X(z) = \sum_n \tau^n \tr_{\bC[X]_n} z.
\end{equation} 
For many examples including $Q=S_{G,g=0,n}\slice{e_1,\ldots,e_n}$, it is known that \begin{equation}
\ch\Higgs(Q)=\SCI_{p=q=0,t=\tau^2}(Q).
\end{equation}

\paragraph{Summary of the functors:}
We summarize below the relationships of the functors discussed so far: 
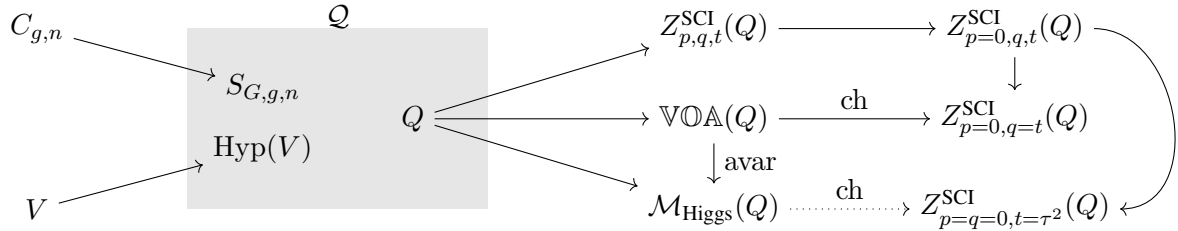
\[
\begin{tikzpicture}[yscale=.8]
\node (V) at (0,0) {$V$};
\node (C) at (0,3) {$C_{g,n}$};
\fill[color=black!10!white] (2,0) rectangle (6,3);
\node at (4,3.2) {$\cQ$};
\node (Hyp) at (3,1) {$\Hyp(V)$};
\node (S) at (3,2) {$S_{G,g,n}$};
\draw[->] (V) to (Hyp);
\draw[->] (C) to (S);
\node (Q) at (5,1.5) {$Q$};
\node (Z) at (9,3) {$\SCI_{p,q,t}(Q)$};
\node (M) at (9,0) {$\Higgs(Q)$};
\node (V) at (9,1.5) {$\VOA(Q)$};
\draw[->] (Q) to (Z);
\draw[->] (Q) to (M);
\draw[->] (Q) to (V);
\draw[->] (V) -- node[midway,right] {$\avar$} (M);
\node (Mac) at (13,3) {$\SCI_{p=0,q,t}(Q)$};
\node (HL) at (13,0) {$\SCI_{p=q=0,t=\tau^2}(Q)$};
\node (Schur) at (13,1.5) {$\SCI_{p=0,q=t}(Q)$};
\draw[->] (Z) to (Mac);
\draw[->,out=0,in=0] (Mac) to (HL);
\draw[->] (Mac) to (Schur);
\draw[->,dotted] (M) -- node[midway,above] {$\ch$} (HL);
\draw[->] (V) -- node[midway,above] {$\ch$} (Schur);
\end{tikzpicture}
\]

\subsection{Consequences}
Let us see a few consequences of the whole setup.

\paragraph{Class S theories of type $\SU(2)$ :}
We already noted that \begin{equation}
S_{\SU(2),0,4}=(T_{\SU(2)}\times T_{\SU(2)})\hypergauge\SU(2)
\end{equation} and also  \begin{equation}
T_{\SU(2)}=\Hyp(V\otimes_\bC V\otimes_\bC V)
\end{equation} where $V$ is the defining 2-dimensional representation of $\SU(2)$.

By applying the functor $\Higgs$, we have \begin{equation}
\Higgs(S_{\SU(2),0,4})=[V_a\otimes_\bC V_b\otimes_\bC V_x\oplus V_y\otimes_\bC V_c\otimes_\bC V_d]\hkq \SU(2).
\end{equation}
Here, we put subscripts to various copies of $V$ to distinguish them, and 
$\SU(2)$ used in the quotient is the diagonal subgroup of $\SU(V_x)$ and $\SU(V_y)$.
The right hand side clearly has an action of $\prod_{i=a,b,c,d}\SU(V_i)$.
But from the left hand side, we see that there should also be an action $S_4$ permuting $\SU(V_{a,b,c,d})$, which is not obvious from the right hand side.
The right hand side, when written as \begin{equation}
V\otimes_\bR \bR^8 \hkq \SU(V),
\end{equation} is the ADHM construction of the minimal nilpotent orbit of $\SO(8)\supset [\prod_{i=a,b,c,d}\SU(V_i)]/\bZ_2$, and the $S_4$ permutations  of $V_i$'s are given by elements of $\Aut(\SO(8))$.

By applying the functor $\SCI_{p,q,t}$, we have the equality \begin{multline}
\SCI(S_{\SU(2),0,4})(a,b,c,d)=\frac{1}{\Gamma_{p,q}(t)\Gamma_{p,q}'(1)}
\frac12\oint\frac{dz}{2\pi \sqrt{-1} z} 
\prod_{\pm}\frac{1}{\Gamma_{p,q}(z^{\pm2})\Gamma_{p,q}(tz^{\pm2})} \\
\times \prod_{\pm\pm\pm}\Gamma_{p,q}(t^{1/2}a^\pm b^\pm z^\pm)
\prod_{\pm\pm\pm}\Gamma_{p,q}(t^{1/2}c^\pm d^\pm z^\pm)
\end{multline}
where  $a$, $b$, $c$, $d$ are now thought of as $\in \U(1)\subset \SU(2)$.
The left hand side should be symmetric under an arbitrary permutation of $a$, $b$, $c$, $d$.
This symmetry is however nontrivial on the right hand side. 
This was pointed out from this physical argument in \cite{Gadde:2009kb},
and completely independently studied and proved in a mathematical work \cite{vdB}.

By applying the functor $\VOA$, we have the equality \begin{equation}
\VOA(S_{\SU(2),0,4}) = \SB[V_a\otimes_\bC V_b\otimes_\bC V_x\oplus V_y\otimes_\bC V_c\otimes_\bC V_d]/\SU(2).
\end{equation}
There is a simple physics argument that the left hand side is just $\hat\so(8)_{-2}$.
Then the equality above is a new free-field construction of this particular vertex operator algebra based on the affine Lie algebra, which remains to be proven.

\paragraph{Class S theories of general type:}
In general, we have the relation \begin{equation}
S_{G,0,4}=(T_{G}\times T_{G})\hypergauge G,
\end{equation}
where $T_G=S_{G,0,3}$ as we defined above.
The left hand side is symmetric under $G^4$ together with $S_4$ permuting four factors of $G$.
The right hand side is symmetric under $G^4$, but only a subgroup $(S_2 \times S_2)\rtimes S_2 \subset S_4$ permuting four factors of $G$ is manifest.

By applying  the functors $\Higgs$ or $\VOA$, we are led to the following conjectures.
Let $X_G=\Higgs(T_{G})$, a hyperk\"ahler cone with $G^3$ action,
and $\bV_G=\VOA(T_{G})$, a vertex operator algebra with a $\hat \fg^3$ subalgebra.
They satisfy the following: 
\begin{itemize}
\item $(X_G \times X_G)\hkq G$ is a hyperk\"ahler cone with $G^4$ action together with $S_4$ permuting four factors of $G$.
\item $(\bV_G\otimes \bV_G)/G$ is a vertex operator algebra with $\hat\fg^{\oplus 4}$ subalgebra with $S_4$ permuting four factors of $\hat \fg$.
\item $X_G=\avar \bV_G$.
\end{itemize}

In addition, there is a way to determine $\SCI_{p=0,q,t}(T_{G})$ explicitly using the theory of Macdonald polynomials. 
By taking a further limit $q=t$ or $q=0$, $t=\tau^2$, we see the properties 
\begin{align}
\ch X_G(z_1,z_2,z_3)&= \sum_\lambda \frac{\prod_{i=1,2,3} K_0(z_i) \underline{H}_\lambda(z_i) }{K_{e_\prin} \underline{H}_\lambda(q^\rho)},&
\ch \bV_G(z_1,z_2,z_3)&= \sum_\lambda \frac{\prod_{i=1,2,3} K_0(z_i) \chi_\lambda(z_i) }{K_{e_\prin} \chi_\lambda(q^\rho)},
\end{align}
where $\lambda$ runs over all irreducible representation of $G$, $\chi_\lambda(z)$ is the  character in that representation, and $\underline{H}_\lambda(z)=N_\lambda H_\lambda(z)$
where $H_\lambda(z)$ is the standard Hall-Littlewood polynomial of type $G$ and $N_\lambda$ is a normalization constant so that $\underline{H}_\lambda$ is  orthonormal under the following measure: \begin{equation}
\delta_{\mu\nu}=\frac1{|W_G|}\int_{T^r} \underline{H}_\lambda(z)\underline{H}_\mu(1/z)
\frac{1}{(1-\tau^2)^r}\prod_{\alpha:\text{roots of $G$}}
\frac{1-z^\alpha}{1-\tau^2 z^\alpha}
\prod_{i=1}^{r}\frac{dz_i}{2\pi \sqrt{-1} z_i} .
\end{equation}

Finally, we should have \begin{align}
X_{\SU(3N)}\slice{[N^3],[N^3],[N^3]}&=\cM^\text{inst}_{E_6,N}, \\
X_{\SU(4N)}\slice{[2N,2N],[N^4],[N^4]}&=\cM^\text{inst}_{E_7,N}, \\
X_{\SU(6N)}\slice{[3N,3N],[2N,2N,2N],[N^6]}&=\cM^\text{inst}_{E_8,N}, 
\end{align} and \begin{align}
\bV_{\SU(3)} &= (\hat{\fe_6})_{-6}, &
\bV_{\SU(4)}\slice{[2,2],[1^4],[1^4]}&=(\hat{\fe_7})_{-8}, &
\bV_{\SU(6)}\slice{[3,3],[2,2,2],[1^6]}&=(\hat{\fe_8})_{-12}.
\end{align}
where the last three equations follow from the property of the Minahan-Nemeschansky theory.

The properties satisfied by $X_G$ were already given in \cite{Moore:2011ee} in a slightly different language,
as a 2d topological QFT taking values in the category of holomorphic symplectic varieties.
Such $X_G$ has now been constructed in \cite{GK,Braverman:2017ofm} as holomorphic symplectic varieties.
The construction of the vertex operator algebras $\bV_G$ satisfying these relations is also announced \cite{ArakawaProc}.

\paragraph{Instantons and W-algebras:}
Finally a brief remark is made about the conjecture that 
the direct sum $\cH_G:=\oplus_n  H_G^*(\cM_{G,n}^\inst)$ of the equivariant cohomology  of the instanton moduli space of simply-laced group $G$ has an action of the W-algebra of the corresponding type,
originally made in \cite{Alday:2009aq}, and now proved by \cite{SchiffmannVasserot,MaulikOkounkov,Braverman:2014xca}, from the point of view of the present note. 

The essential point is that there is an another functor $Z_\text{Nek}$ defined on $\cQ_c(G)$ taking values in $\cH_G$, such that a family of objects in $\cQ_c(G)$ parameterized by $\sM$ is sent to a section of an $\cH_G$ bundle over $\sM$.
When $\sM=\sM_{g,n}$ as in the class S theory, this has a natural relationship with the theory of the 2d conformal blocks.
A consideration in this line of thought naturally leads to the conjecture that $\cH_G$ has to have the action of the W-algebra of type $G$.
More details can be found in the longer version of this article on the author's webpage.

\bibliographystyle{hyperamsalpha}
\let\bbb\bibitem\def\bibitem{\itemsep1pt\bbb}
\bibliography{boo}

\end{document}